\documentclass[aps,prl,twocolumn,superscriptaddress]{revtex4}
\usepackage{graphicx}
\usepackage{amssymb}
\usepackage{pifont}

\begin{document}



\title{Strong-coupling Spin-singlet Superconductivity with Multiple Full Gaps in Hole-doped Ba$_{0.6}$K$_{0.4}$Fe$_2$As$_2$ Probed by $^{57}$Fe-NMR}

\author{M.~Yashima}
\affiliation{Graduate School of Engineering Science, Osaka University, Osaka 560-8531, Japan}
\affiliation{JST, TRiP, 1-2-1 Sengen, Tsukuba 305-0047, Japan}
\author{H.~Nishimura}
\affiliation{Graduate School of Engineering Science, Osaka University, Osaka 560-8531, Japan}
\author{H.~Mukuda}
\affiliation{Graduate School of Engineering Science, Osaka University, Osaka 560-8531, Japan}
\affiliation{JST, TRiP, 1-2-1 Sengen, Tsukuba 305-0047, Japan}
\author{Y.~Kitaoka}
\affiliation{Graduate School of Engineering Science, Osaka University, Osaka 560-8531, Japan}
\author{K.~Miyazawa}
\affiliation{National Institute of Advanced Industrial Science and Technology (AIST), Tsukuba, 305-8568, Japan}
\author{P.~M.~Shirage}
\affiliation{National Institute of Advanced Industrial Science and Technology (AIST), Tsukuba, 305-8568, Japan}
\author{K.~Kiho}
\affiliation{National Institute of Advanced Industrial Science and Technology (AIST), Tsukuba, 305-8568, Japan}
\author{H.~Kito}
\affiliation{National Institute of Advanced Industrial Science and Technology (AIST), Tsukuba, 305-8568, Japan}
\affiliation{JST, TRiP, 1-2-1 Sengen, Tsukuba 305-0047, Japan}
\author{H.~Eisaki}
\affiliation{National Institute of Advanced Industrial Science and Technology (AIST), Tsukuba, 305-8568, Japan}
\affiliation{JST, TRiP, 1-2-1 Sengen, Tsukuba 305-0047, Japan}
\author{A.~Iyo}
\affiliation{National Institute of Advanced Industrial Science and Technology (AIST), Tsukuba, 305-8568, Japan}
\affiliation{JST, TRiP, 1-2-1 Sengen, Tsukuba 305-0047, Japan}

\begin{abstract}
We present $^{57}$Fe-NMR measurements of the novel normal and superconducting-state characteristics of the iron-arsenide superconductor Ba$_{0.6}$K$_{0.4}$Fe$_2$As$_2$ ($T_c$ = 38 K). In the normal state, the measured Knight shift and nuclear spin-lattice relaxation rate $(1/T_1)$ demonstrate the development of wave-number ($q$)-dependent spin fluctuations, except at $q$ = 0, which may originate from the nesting across the disconnected Fermi surfaces. In the superconducting state, the spin component in the $^{57}$Fe-Knight shift decreases to almost zero at low temperatures, evidencing a spin-singlet superconducting state. The $^{57}$Fe-$1/T_1$ results are totally consistent with a $s^\pm$-wave model with multiple full gaps, regardless of doping with either electrons or holes.
\end{abstract}
\vspace*{5mm}

\maketitle

The recent discovery of superconductivity (SC) in the iron (Fe)-based oxypnictide LaFeAsO$_{1-x}$F$_x$ at the SC transition temperature $T_c=26$  K has provided a new route toward the realization of high-$T_c$ SC \cite{Kamihara}. The mother material, LaFeAsO, exhibits a structural phase transition from tetragonal (P4/nmm) to orthorhombic (Cmma) form at $T \sim$ 155 K and then exhibits a striped antiferromagnetic (AFM) order with ${\bf Q}$ = (0, $\pi$) or ($\pi$, 0) and $T_N \sim$ 140 K \cite{Cruz}. The calculated Fermi surfaces (FSs) for undoped LaFeAsO consist of two small electron cylinders around the tetragonal M point and two hole cylinders, plus a heavy 3D hole pocket, around the $\Gamma$ point \cite{Singh}. Measurements of the nuclear spin-lattice relaxation rate ($1/T_1$) for the LaFeAsO system in the SC state revealed the lack of a coherence peak below $T_c$ and the presence of $T^3$-like behavior, suggesting an unconventional SC nature with line-node gaps \cite{Nakai,Grafe,Mukuda}. However, some experiments to measure parameters such as penetration depth, together with studies using angle resolved photoemission spectroscopy (ARPES), have shown that the SC order parameter (OP) is of fully gapped $s$-wave symmetry \cite{Luetkens,Ding,Hashimoto,Kondo}. The theory was the first to propose $s^{\pm}$-wave pairing symmetry as a promising candidate for the SC state in Fe-pnictide superconductors \cite{Mazin,Kuroki}.

\begin{figure}[htbp]
\centering
\includegraphics[width=6.5cm]{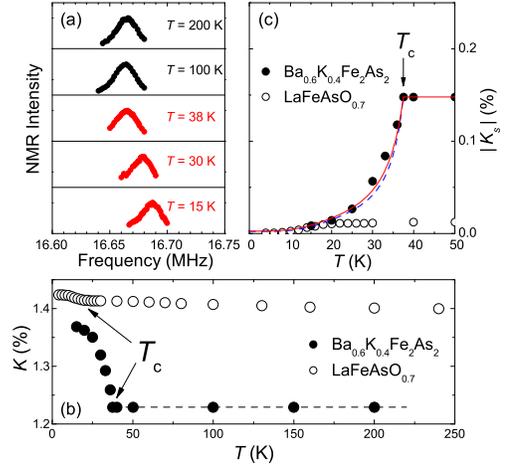}
\caption[]{\footnotesize (Color online) $T$ dependence of (a) $^{57}$Fe-NMR spectrum and (b) Knight shift ($^{57}K$) at $H$ = 11.966 T parallel to the $ab$ plane in Ba$_{0.6}$K$_{0.4}$Fe$_2$As$_2$. (c) The spin component $^{57}K_s$ in $^{57}K$ deduced by subtracting $K_{orb}\sim$ 1.38 \%, revealing a spin-singlet SC state similar to that in LaFeAsO$_{0.7}$ \cite{Terasaki}. The solid and dashed lines indicate the calculated results assuming Model A and Model B with the same fitting parameters as those in $1/T_1$, respectively (see text). 
}
\label{Knight}
\end{figure}

Another family of FeAs-based superconductors without oxygen has been reported in hole-doped Ba$_{1-x}$K$_x$Fe$_2$As$_2$ with $T_c=38$ K \cite{Rotter1}. The mother material, BaFe$_2$As$_2$, has a ThCr$_2$Si$_2$-type structure and consists of alternating layers comprising FeAs$_4$ tetrahedra and Ba. BaFe$_2$As$_2$ also exhibits a structural phase transition from the tetragonal (I4/mmm) to orthorhombic (Fmmm) form, accompanied by a striped AFM order with ${\bf Q}$ = ($\pi$, 0, $\pi$) at $T_N$ = 140 K \cite{Rotter2,Huang}. The previous $^{75}$As-NMR study on Ba$_{0.6}$K$_{0.4}$Fe$_2$As$_2$ reported that $1/T_1$ shows $T^3$-like behavior well below $T_c$ \cite{Fukazawa,MukudaPhysC}, which is in contrast to the fully gapped $s$-wave symmetry of the SC OP, as revealed by ARPES on hole-doped Ba$_{0.6}$K$_{0.4}$Fe$_2$As$_2$ and electron-doped Ba(Fe$_{0.85}$Co$_{0.15}$)$_2$As$_2$ \cite{Ding,Terashima}. Furthermore, ARPES results for Ba$_{0.6}$K$_{0.4}$Fe$_2$As$_2$ show that there are two SC gaps with different values: a large gap on the two, small, hole-like and electron-like FS sheets, and a small gap on the large, hole-like FS \cite{Ding}. This two-SC-gap phenomenon was suggested by NMR studies as well \cite{Kawasaki,Matano}. In this letter, we report the results of microscopic $^{57}$Fe-NMR measurements on hole-doped Ba$_{0.6}$K$_{0.4}$Fe$_2$As$_2$ with $T_c=38$ K enriched with the isotope $^{57}$Fe to address its normal-state and SC characteristics.

\begin{figure}[htbp]
\centering
\includegraphics[width=5.5cm]{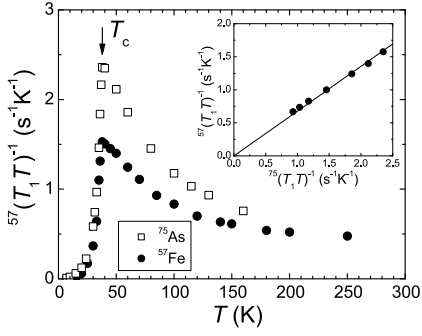}
\caption[]{\footnotesize  (Color online) $T$ dependence of $^{57}$Fe-$(T_1T)^{-1}$ in Ba$_{0.6}$K$_{0.4}$Fe$_2$As$_2$, along with that in $^{75}$As-$(T_1T)^{-1}$. The inset shows the plot of $^{57}$Fe-$(T_1T)^{-1}$ against $^{75}$As-$(T_1T)^{-1}$ with the implicit parameter $T$; $^{57}(T_1T)^{-1}/^{75}(T_1T)^{-1}\sim$ 0.68. 
}
\label{iT1T}
\end{figure}

A polycrystalline sample of $^{57}$Fe-enriched Ba$_{0.6}$K$_{0.4}$Fe$_2$As$_2$ was synthesized by the high-pressure synthesis technique, as described elsewhere \cite{Shirage}. Powder X-ray diffraction measurements indicated that the Ba$_{0.6}$K$_{0.4}$Fe$_2$As$_2$ sample almost completely consisted of a single phase with the lattice parameters $a$ = 3.9142 \AA and $c$ = 13.305 \AA. The sample was moderately crushed into powder for the NMR measurements, which were easily performed under a strong magnetic field along the direction including the $ab$ plane. The $^{57}$Fe- and $^{75}$As-NMR measurements were performed on a phase coherent pulsed NMR spectrometer at respective magnetic fields $H$ of 11.966 and 5.12 T. $T_1$ was measured with a saturation recovery method.

Figure 1(a) shows the $T$ dependence of the $^{57}$Fe-NMR spectra at $H$ = 11.966 T parallel to the $ab$ plane. The Knight shift $^{57}K^{ab}$ stays almost constant above $T_c$, followed by a steep increase upon cooling below $T_c$, as shown in Fig. 1(b). The Knight shift comprises a spin component and an orbital component, denoted as $K_s$ and $K_{orb}$, respectively. Note that $K_s=A_{hf}\chi_s(T)$ depends on $T$, but $K_{orb}$ does not. Here, $A_{hf}$ is the hyperfine coupling constant and $\chi_s(T)$ is the uniform spin susceptibility. Since $^{57}A^{ab}_{hf}$ is known to be negative due to the inner core-polarization effect, as in LaFeAsO$_{0.7}$ \cite{Terasaki}, the increase in $^{57}K^{ab}$ below $T_c$ (see Fig. 1(b)) indicates the decrease in $\chi^{ab}_s(T)$, demonstrating the formation of a spin-singlet SC state, as in LaFeAsO$_{0.7}$. If we assume $^{57}K^{ab}_{orb}\sim$ 1.38 \% in this compound, $^{57}K^{ab}_{s}$ goes down to zero at $T$ = 0, as displayed in Fig. 1(c).

\begin{figure}[htbp]
\centering
\includegraphics[width=5cm]{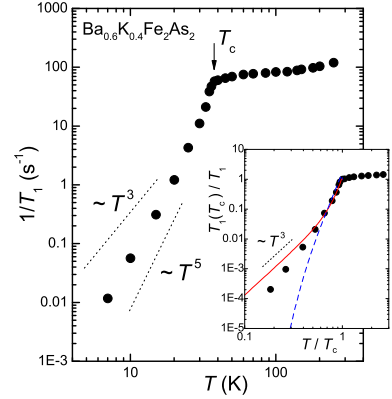}
\caption[]{\footnotesize  (Color online) $T$ dependence of $^{57}(1/T_1)$ for Ba$_{0.6}$K$_{0.4}$Fe$_2$As$_2$. The $T$ dependence of $^{57}(1/T_1)$ exhibits a $T^5$-like dependence well below $T_c$, which cannot be simulated with either a simple $d_{x^2-y^2}$-wave model (solid curve) or an isotropic $s$-wave model with no coherence effect (dashed curve; see inset).}
\label{iT1}
\end{figure}

$^{57}$Fe-$1/T_1$ was uniquely determined with a single component throughout the whole $T$ range. A large enhancement of $^{57}$Fe-$(T_1T)^{-1}$ is observed on cooling down to $T_c$ from the normal state, as also seen in the results for $^{75}$As-$(T_1T)^{-1}$ (Fig. 2). By contrast, note that $^{57}K^{ab}_{s}$, which is in proportion to $\chi_s(q=0, \omega=0)$, stays almost constant above $T_c$. In general, $(T_1T)^{-1}$ is expressed as  
\[
\frac{1}{T_1T} = \frac{\pi k_B \gamma_n^2}{(\gamma_e \hbar)^2} \sum_{\textit{\textbf{q}}}A_{hf}^2(\textit{\textbf{q}}) \frac{\chi''_{\bot}(\textit{\textbf{q}},\omega_0)}{\omega_0},
\]
where $\chi''_{\bot}(\textit{\textbf{q}},\omega_0)$ is the imaginary part of the dynamical susceptibility in a direction perpendicular to the applied magnetic field, $\omega_0$ is the NMR frequency, and $\gamma_n$ is the nuclear gyromagnetic ratio. The difference in $T$ dependency between the Knight shift and $(T_1T)^{-1}$ points to the development of $q$-dependent spin fluctuations, except at $q$ = 0, upon cooling. To unravel the characteristics of these spin fluctuations, compare the $(T_1T)^{-1}$ values at Fe and As sites; $^{57}$Fe-$(T_1T)^{-1}/^{75}$As-$(T_1T)^{-1}\sim$ 0.68 and is almost constant in the range $T_c$(= 38 K) -- 100 K, as shown in the inset of Fig. 2. If ferromagnetic spin fluctuations are predominant around $q$ = 0, the ratio $^{57}(T_1T)^{-1}/^{75}(T_1T)^{-1} \simeq$ 0.038 is estimated using $(^{57}A_{hf}/^{75}A_{hf})^2_{q=0} \sim$ 1. This value is about one order of magnitude smaller than the experimental value, indicating that ferromagnetic spin fluctuations are not developed. Furthermore, according to the argument in the literature \cite{Terasaki}, spin fluctuations around ${\bf Q}=(\pi,\pi)$ are not responsible for the large enhancement in $(T_1T)^{-1}$ values at both the Fe and As sites. Given these facts, it is likely that the enhancement is the results of the spin fluctuations with ${\bf Q}=(\pi,0)$ and $(0,\pi)$ that would be expected from the interband nesting.

\begin{figure}[htbp]
\centering
\includegraphics[width=7.5cm]{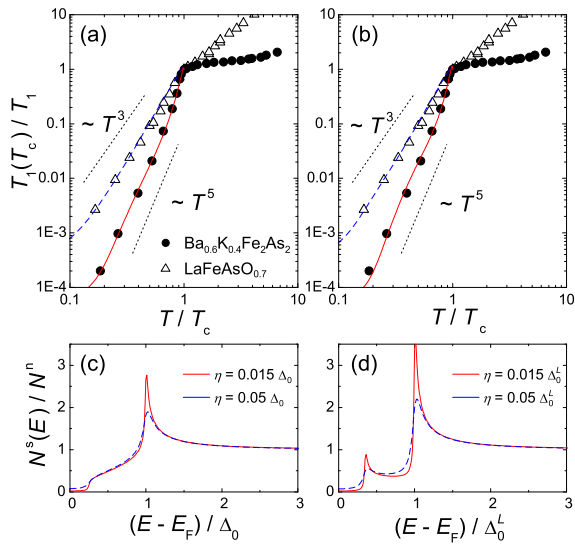}
\caption[]{\footnotesize  (Color online) The simulations are based on (a) Model A and (b) Model B. The experimental $^{57}(1/T_1)$ data for Ba$_{0.6}$K$_{0.4}$Fe$_2$As$_2$ and LaFeAsO$_{0.7}$ \cite{Terasaki} are normalized by the value at $T_c$. Either model can qualitatively reproduce the experimental results, assuming reasonable fitting parameters (see Table I). The energy dependences of the density of states used in these simulations are based on (c) Model A and (d) Model B.
}
\label{DOS}
\end{figure}

Here, we remark that on the basis of the fluctuation-exchange approximation (FLEX) on an effective five-band Hubbard model, the recent theoretical work \cite{Ikeda} appears to explain qualitatively the evolution of magnetic characteristics from that in electron-doped systems to that in hole-doped systems: in electron-doped systems, $(T_1T)^{-1}$ and spin susceptibility decrease significantly upon cooling, producing a pseudogap behavior that originates from the band-structure effect, i.e., the existence of a high density of states just below the Fermi level. In hole-doped systems, $(T_1T)^{-1}$ is enhanced upon cooling due to the nesting across the disconnected FSs, but the Knight shift is not. In this model, it was suggested that the dominant contribution to the pairing interaction for $s^{\pm}$-wave SC comes from an interplay between spin-dipole and spin-quadrupole and not from striped AFM spin fluctuations at ${\bf Q}=(\pi,0)$ and $(0,\pi)$.

Next, we discuss the SC characteristics of Ba$_{0.6}$K$_{0.4}$Fe$_2$As$_2$ through the $T$ dependence of $^{57}$Fe-$1/T_1$, as shown in Fig. 3. In the SC state, $^{57}$Fe-$1/T_1$ steeply decreases upon cooling, without a coherence peak just below $T_c$, pointing to the unconventional SC nature of this compound and electron-doped LaFeAsO systems \cite{Terasaki}. On the other hand, ARPES revealed nearly isotropic and nodeless SC gaps with different values on three electron and hole FSs \cite{Ding}. Motivated by these seemingly incompatible experimental results, an extended $s^{\pm}$-wave model with a sign reversal of the OP among the FSs has been proposed \cite{Mazin,Kuroki,Nagai,Ikeda}. Note that the $T$ dependence of $^{57}$Fe-$(1/T_1)$ below $T_c$ cannot be simulated with either a simple $d_{x^2-y^2}$-wave SC model($\Delta(\phi)=\Delta_0 \cos 2\phi$ with $2\Delta_0/k_BT_c=10$) or an isotropic $s$-wave model($2\Delta_0/k_BT_c=8$) with no coherence effect, as shown in the inset of Fig. 3. Therefore, we have applied a multiple SC gap model to interpret the $T$ dependence of $^{57}$Fe-$(1/T_1)$ below $T_c$.

\begin{table}
\centering
\caption[]{\footnotesize The sizes of the SC gaps, $N_{FS1}/(N_{FS1}+N_{FS2})$, and $\eta/\Delta_0$ estimated by assuming Model A and Model B for Ba$_{0.6}$K$_{0.4}$Fe$_2$As$_2$ and LaFeAsO$_{0.7}$.}

\begin{tabular}{lccc}
\hline
 Model & Fitting & Ba$_{0.6}$K$_{0.4}$Fe$_2$As$_2$ & LaFeAsO$_{0.7}$ \\
  & parameters & $T_c$ = 38 K & $T_c^{H=0}$ = 28 K \\
\hline
\quad A & $2\Delta_0/k_BT_c$ & 9.2 & 4.6 \\
  & $\Delta_{min}/\Delta_0$ & 0.25  & 0.25 \\
  & $\frac{N_{FS1}}{N_{FS1}+N_{FS2}}$ & 0.4  & 0.4 \\
  & $\eta/\Delta_0$& 0.015 & 0.05 \\
\hline
\quad B & $2\Delta_0^L/k_BT_c$ & 9.4 & 4.4 \\
  & $\Delta_0^S/\Delta_0^L$ & 0.35 & 0.35 \\
  & $\frac{N_{FS1}}{N_{FS1}+N_{FS2}}$ & 0.7  & 0.7 \\
  & $\eta/\Delta_0^L$ & 0.015  & 0.05 \\
\hline
\end{tabular}
\label{table1}
\end{table}

First, we proceed with an analysis of $1/T_1$ by applying an anisotropic two-full-gap $s^\pm$-wave model in which one of the gaps is anisotropic and the other isotropic, denoted as Model A. This was proposed to explain the $T^3$ behavior of $1/T_1$ observed in LaFeAsO$_{0.6}$ on the basis of the effective five-band model \cite{Nagai}. For simplification, we assume that there are two FSs (FS1 and FS2) dominated by an isotropic full gap and an anisotropic full gap in the SC state, respectively, and neglect the coherence factor. These gap functions are given by $\Delta^{FS1}= \Delta_0$ and $\Delta^{FS2}(\phi)=(\Delta_0+\Delta_{min})/2 + (\Delta_0-\Delta_{min})\cos 2\phi/2$. Here $\Delta_{min}$ gives rise to an anisotropy in a full gap function, as described in Ref. \cite{Nagai}. According to Nagai et al., when we assume a fraction of the density of states (DOS) at FS1, $N_{FS1}/(N_{FS1}+N_{FS2})$ = 0.4, the experiment becomes consistent with a calculation using the parameters $\Delta_{min}$ = 0.25 $\Delta_0$, $2\Delta_0/k_BT_c$ = 9.2, and the smearing factor $\eta$ = 0.015 $\Delta_0$ for the DOS (to compensate for impurity scattering), as shown by solid line in Fig. 4(a). Here, $N_{FS1}$ and $N_{FS2}$ are the respective DOSs at  FS1 and FS2. We state that this model reproduced the $^{57}(1/T_1)$ result for LaFeAsO$_{0.7}$ using parameters $\Delta_{min}$ = 0.25 $\Delta_0$, $2\Delta_0/k_BT_c$ = 4.6, and $\eta$ = 0.05 $\Delta_0$, which is also shown by the dashed curve in Fig. 4(a). The $2\Delta_0/k_BT_c$ = 9.2 in Ba$_{0.6}$K$_{0.4}$Fe$_2$As$_2$ is twice that in LaFeAsO$_{0.7}$ (4.6), which reveals that a strong-coupling SC state is realized in Ba$_{0.6}$K$_{0.4}$Fe$_2$As$_2$, thus increasing $T_c$. The fact that $\eta$ in Ba$_{0.6}$K$_{0.4}$Fe$_2$As$_2$ is smaller than that in LaFeAsO$_{0.7}$ ($\eta^{Ba122}/\eta^{La1111} \simeq$ 0.3) implies that the SC in Ba$_{0.6}$K$_{0.4}$Fe$_2$As$_2$ is more robust to impurity scattering because of the larger SC gap $2\Delta_0$. In Table \ref{table1}, we have summarized the experimentally obtained parameters for Ba$_{0.6}$K$_{0.4}$Fe$_2$As$_2$ and LaFeAsO$_{0.7}$. The important outcome is that the respective $1/T_1$ results for Ba$_{0.6}$K$_{0.4}$Fe$_2$As$_2$ and LaFeAsO$_{0.7}$, which seemingly follow a $T^5$- and a $T^3$- like behaviors below $T_c$, are consistently explained in terms of Model A only by changing the size of the SC gap $2\Delta_0$.

Next, since the ARPES experiment on Ba$_{0.6}$K$_{0.4}$Fe$_2$As$_2$ revealed that the anisotropy of the gap is small in every FS sheet \cite{Ding}, we tentatively apply an isotropic two-full-gap $s^\pm$-wave model in which both gaps are isotropic, denoted as Model B. Assuming that the large ($\Delta_0^{L}$) and small ($\Delta_0^{S}$) isotopic full gaps open on FS1 and FS2, respectively, a good fitting to the experiment is possible as shown by the solid line in Fig. 4(b) using parameters $2\Delta_0^{L}/k_BT_c=9.4$, $\Delta_0^{S}/\Delta_0^{L}$ = 0.35, $N_{FS1}/(N_{FS1}+N_{FS2})$ = 0.7, and $\eta$ = 0.015 $\Delta_0^{L}$. However, the experiment does not replicate the slight step-wise behavior predicted by the calculation. Remarkably, the simulated result for the gap ratio $\Delta_0^{S}/\Delta_0^{L}$ = 0.35 is comparable to the value estimated by ARPES ($\sim$0.44) \cite{Ding}. As for LaFeAsO$_{0.7}$, we note that the $1/T_1$ data are also reproduced as shown by the dashed line in Fig. 4(b), using a smaller gap $2\Delta_0^{L}/k_BT_c$ = 4.4 and a larger $\eta$ = 0.05 $\Delta_0^{L}$ than those in Ba$_{0.6}$K$_{0.4}$Fe$_2$As$_2$. In this context, we cannot rule out at the present stage that Model B also explains both the experimental results in Ba$_{0.6}$K$_{0.4}$Fe$_2$As$_2$ and LaFeAsO$_{0.7}$. In order to shed light on the differences between the two models, the quasiparticle DOSs in the SC state are shown in Figs. 4(c) and 4(d), respectively. To identify the more appropriate model, further precise measurements are required on the hole-doped systems, involving systematic variation in the hole-doping level.

In conclusion, $^{57}$Fe-NMR studies on the hole-doped Ba$_{0.6}$K$_{0.4}$Fe$_2$As$_2$ with $T_c=38$ K have unraveled novel normal- and SC-state characteristics. Spin fluctuations with finite $q$-vectors develop upon cooling down to $T_c$; they may originate from the nesting across the disconnected FSs with ${\bf Q}$ = ($\pi$, 0) and (0, $\pi$). The $^{57}$Fe-$(1/T_1)$ results have revealed that Model A is consistently applicable not only to hole-doped Ba$_{0.6}$K$_{0.4}$Fe$_2$As$_2$ with $T_c$ = 38 K but also to electron-doped LaFeAsO$_{0.7}$ with $T_c$ = 24 K. But Model B cannot be ruled out in understanding the present Fe-NMR results. In any case, the $2\Delta_0/k_BT_c$ value of Ba$_{0.6}$K$_{0.4}$Fe$_2$As$_2$ is almost twice that of LaFeAsO$_{0.7}$, which reveals that a strong coupling SC state is realized in Ba$_{0.6}$K$_{0.4}$Fe$_2$As$_2$. We suggest that the pairing mechanism of the unconventional SC with multiple fully gapped $s^{\pm}$-wave symmetry may be universal to the Fe-based superconductors.

This work was supported by a Grant-in-Aid for Specially Promoted Research (20001004) and partially supported by the Global COE Program (Core Research and Engineering of Advanced Materials-Interdisciplinary Education Center for Materials Science) from the Ministry of Education, Culture, Sports, Science and Technology (MEXT), Japan. M. Y. was supported by a Grant-in-Aid for Young Scientists (B) of MEXT (20740175).

\end{document}